\documentclass[12pt,preprint]{aastex}

\newcommand{\dechms}[4]{$#1^{\rm h}#2^{\rm m}#3\mbox{$^{\rm s}\mskip-7.6mu.\,$}#4$}

\newcommand{\decdms}[4]{$#1^{\circ}#2'#3\mbox{$''\mskip-7.6mu.\,$}#4$}
\newcommand{\msec}[2]{$#1\mbox{$''\mskip-7.6mu.\,$}#2$}
\newcommand{\mmsec}[2]{$#1\mbox{$^s\mskip-7.6mu.\,$}#2$}
\newcommand{\mdeg}[2]{$#1\mbox{$^\circ \mskip-7.6mu.\,$}#2$}

\newcommand{\Msun}{M$_{\odot}$}
\newcommand{\Rsun}{R$_{\odot}$}

\begin{document}

\title{VLBA determination of the distance to nearby star-forming regions\\
I. The distance to T Tauri with 0.4\% accuracy}

\author{Laurent Loinard, Rosa M.\ Torres}
\affil{Centro de Radiostronom\'{\i}a y Astrof\'{\i}sica, 
       Universidad Nacional Aut\'onoma de M\'exico,\\
       Apartado Postal 72--3 (Xangari), 58089 Morelia, Michoac\'an, M\'exico;\\       l.loinard@astrosmo.unam.mx}

\author{Amy J.\ Mioduszewski}
\affil{National
    Radio Astronomy Observatory, Array Operations Center,\\ 1003
    Lopezville Road, Socorro, NM 87801, USA}

\author{Luis F.\ Rodr\'{\i}guez, Rosa A.\ Gonz\'alez-L\'opezlira, R\'egis Lachaume}
\affil{Centro de Radiostronom\'{\i}a y Astrof\'{\i}sica,
       Universidad Nacional Aut\'onoma de M\'exico,\\
       Apartado Postal 72--3 (Xangari), 58089 Morelia, Michoac\'an, M\'exico}

\and

\author{Virgilio V\'azquez \& Erandy Gonz\'alez}
\affil{Universidad Tecnol\'ogica de la Mixteca, Carretera Huajuapan-Acatlima\\ 
       69000 Huajuapan de Le\'on, Oaxaca, M\'exico}

\begin{abstract}

In this article, we present the results of a series of twelve 3.6-cm
radio continuum observations of T Tau Sb, one of the companions of the
famous young stellar object T Tauri. The data were collected roughly
every two months between September 2003 and July 2005 with the Very
Long Baseline Array (VLBA). Thanks to the remarkably accurate
astrometry delivered by the VLBA, the absolute position of T Tau Sb
could be measured with a precision typically better than about 100
micro-arcseconds at each of the twelve observed epochs. The trajectory
of T Tau Sb on the plane of the sky could, therefore, be traced very
precisely, and modeled as the superposition of the trigonometric
parallax of the source and an accelerated proper motion.  The best fit
yields a distance to T Tau Sb of 147.6 $\pm$ 0.6 pc.  The observed
positions of T Tau Sb are in good agreement with recent infrared
measurements, but seem to favor a somewhat longer orbital period than
that recently reported by Duch\^ene et al. (2006) for the T Tau Sa/T
Tau Sb system.

\end{abstract}

\keywords{Astrometry --- Stars: individual (T Tau) --- Radiation
  mechanisms: non-thermal --- Magnetic fields --- stars: formation ---
  Binaries: general}

\section{Introduction} 

To provide accurate observational constraints for pre-main sequence
evolutionary models, and thereby improve our understanding of
star-formation, it is crucial to measure as accurately as possible the
properties (age, mass, luminosity, etc.) of individual young stars.
The determination of most of these parameters, however, depends
critically on the often poorly known distance to the object under
consideration. While the average distance to nearby low-mass star-forming
regions 
(e.g.\ Taurus or $\rho-$Ophiuchus) has been estimated to about 20\%
precision using indirect methods (Elias 1978a,b, Kenyon et al.\ 1994,
Knude \& Hog 1998), the line-of-sight depth of these regions is
largely unknown, and accurate distances to individual objects are
still missing.  Even the highly successful Hipparchos mission
(Perryman et al.\ 1997) did little to improve the situation (Bertout
et al.\ 1999) because young stars are still heavily embedded in their
parental clouds and are, therefore, faint in the optical bands
observed by Hipparchos. Future space missions such as GAIA will
undoubtedly have the capacity to accurately measure the trigonometric
parallax of optically fainter stars, but these missions will still be
unable to access very deeply embedded sources, and will only start to
provide results in about a decade. In the meantime, extremely high
quality infrared and X-ray surveys of many star-forming regions are
being obtained (e.g.\ Evans et al.\ 2003, G\"udel et al.\ 2007), and
their potential cannot be fully exploited because of the
unavailability of good distance estimates.

Low-mass young stars often generate non-thermal continuum emission
produced by the interaction of free electrons with the intense
magnetic fields that tend to exist near their surfaces (e.g.\
Feigelson \& Montmerle 1999). Since the magnetic field strength
decreases quickly with the distance to the stellar surface (as
$r^{-3}$ in the magnetic dipole approximation), the emission is
strongly concentrated to the inner few stellar radii. If the magnetic
field intensity and the electron energy are sufficient, the resulting
compact radio emission can be detected with Very Long Baseline
Interferometers (VLBI --e.g.\ Andr\'e et al.\ 1992). The relatively
recent possibility of accurately calibrating the phase of VLBI
observations of faint, compact radio sources using nearby quasars
makes it possible to measure the absolute position of these objects
(or, more precisely, the angular offset between them and the
calibrating quasar) to better than a tenth of a milli-arcsecond
(Loinard et al.\ 2005, see also below). This level of precision is
sufficient to constrain the trigonometric parallax of sources within a
few hundred parsecs of the Sun (in particular of nearby young stars)
with a precision better than a few percents using multi-epoch VLBI
observations.

Taking advantage of this situation, we have recently initiated a large
project aimed at accurately measuring the trigonometric parallax of a
significant sample of magnetically active young stars in nearby
star-forming regions (Taurus, $\rho-$Ophiuchus, Perseus, Serpens, and
Cepheus) using the 10-element Very Long Baseline Array (VLBA) of the
National Radio Astronomy Observatory (NRAO). In the present article,
we will concentrate on T Tau Sb, one of the members of the famous
young stellar system T Tauri (see e.g.\ Duch\^ene et al.\ 2006 for a
recent summary of the properties of that system). T Tau Sb has long
been known to be associated with a compact non-thermal radio source
(Skinner \& Brown 1994; Phillips et al.\ 1993, Johnston et al.\ 2003)
characterized by strong variability and significant circular
polarization. An extended, thermal, radio halo studied in detail by
Loinard et al.\ (2007) and probably related to stellar winds, also
exist around T Tau Sb. While this extended structure contributes to
the total radio flux as measured, for instance, with the VLA, it is
effectively filtered out in Very Long Baseline Interferometry
experiments. Indeed, in the intercontinental VLBI observations
published by Smith et al.\ (2003), only about 40\% of the
simultaneously measured VLA flux density is retrieved. The radio
source detected by Smith et al.\ (2003) is very compact (R $<$ 15
\Rsun), and its flux was about 3 mJy at the time of their
observations. Its trajectory over the plane of the sky was studied by
Loinard et al.\ (2005) using a series of 7 VLBA
observations. Unfortunately, these data were recently found to have
been affected by a bug that caused the VLBA correlator to use
predicted rather than measured Earth Orientation Parameters (see
http://www.vlba.nrao.edu/astro/messages/eop/). This problem corrupted
the visibility phases, and strongly affected the quality of the
astrometry of the data published in Loinard et al.\ (2005). The
post-fit rms for the data published by Loinard et al.\ (2005) was
about 250 mas compared with 60--90 mas for the present data (see
below). Here, we will re-analyze these VLBA data, and combine them
with 5 newer observations to measure the trigonometric parallax, and
study the proper motion of T Tau Sb.

\section{Observations and data calibration}

In this paper, we will make use of a series of twelve continuum 3.6 cm
(8.42 GHz) observations of T Tau Sb obtained every two months between
September 2003 and July 2005 with the VLBA (Tab.\ 1). Our phase center
was at $\alpha_{J2000.0}$ = \dechms{04}{21}{59}{4263},
$\delta_{J2000.0}$ = +\decdms{19}{32}{05}{730}, the position of the
compact source detected by Smith et al.\ (2003). Each observation
consisted of series of cycles with two minutes spent on source, and
one minute spent on the main phase-referencing quasar J0428+1732,
located \mdeg{2}{6} away.  J0428+1732 is a very compact extragalactic
source whose absolute position ($\alpha_{J2000.0}$ = \dechms{04}{28}{35}{633679},$\delta_{J2000.0}$ = \decdms{17}{32}{23}{58799}) is known to better than 1
milli-arcsecond ($\sigma_\alpha$ = 0.59 mas, $\sigma_\delta$ = 0.89
mas; Beasley et al.\ 2002). During the first 6
observations, the secondary quasar J0431+1731 was also observed
periodically --about every 30 minutes-- to check the astrometric
quality of the data, and to compare our results with those of Smith et
al.\ (2003) who used J0431+1731 as their phase calibrator. A detailed
comparison with the results of Smith et al.\ (2003) and with the
numerous VLA observations available from the literature, however, will
be postponed to a forthcoming article.

The data were edited and calibrated using the Astronomical Image
Processing System (AIPS --Greisen 2003). The basic data reduction 
followed the
standard VLBA procedures for phase-referenced observations. First, the
most accurate measured Earth Orientation Parameters obtained from the
US Naval Observatory database were applied to the data in order to
correct the erroneous values initially used by the VLBA
correlator. Second, dispersive delays caused by free electrons in the
Earth's atmosphere were accounted for using estimate of the electron
content of the ionosphere derived from Global Positioning System (GPS)
measurements. {\it A priori} amplitude calibration based on the
measured system temperatures and standard gain curves was then
applied. The fourth step was to correct the phases for antenna
parallactic angle effects, and the fifth was to remove residual
instrumental delays caused by the VLBA electronics. This was done by
measuring the delays and phase residuals for each antenna and IF using
the fringes obtained on a strong calibrator. The final step of this
initial calibration was to remove global frequency- and time-dependent
phase errors using a global fringe fitting procedure on the main phase
calibrator (J0428+1732), which was assumed at this stage to be a point
source.

In this initial calibration, the solutions from the global fringe fit
were only applied to the main phase calibrator itself. The
corresponding calibrated visibilities were then imaged, and several
passes of self-calibration were performed to improve the overall
amplitude and phase calibration. In the image obtained after the
self-calibration iterations, the main phase calibrator is found to be
slightly extended. To take this into account, the final global fringe
fitting part of the reduction was repeated using the image of the main
phase calibrator as a model instead of assuming it to be a point
source. Note that a different phase calibrator model was produced for
each epoch to account for possible small changes in the main
calibrator structure from epoch to epoch. The solutions obtained after
repeating this final step were edited for bad points and applied to
the target source.  Using an image model for the calibrator
rather than assuming a point source improved the position 
accuracy by a few tens of $\mu$as.

Because of the significant overheads that were necessary to properly
calibrate the data, only about 3 of the 6 hours of telescope time
allocated to each of our observations were actually spent on
source. Once calibrated, the visibilities were imaged with a pixel
size of 50 $\mu$as after weights intermediate between natural and
uniform (ROBUST = 0 in AIPS) were applied. This resulted in a typical
r.m.s.\ noise level of 70 $\mu$Jy for most observations, though for a
few epochs with less favorable weather conditions, the noise
level exceeded 100 $\mu$Jy (Tab.\ 1). T Tau Sb was detected with a
signal to noise better than 10 at each epoch (Tab.\ 1), and its
absolute position (listed in columns 2 and 4 of Tab.\ 1) was
determined using a 2D Gaussian fitting procedure (task JMFIT in
AIPS). This task provides an estimate of the position error (columns 3
and 5 of Tab.\ 1) based on the expected theoretical astrometric
precision of an interferometer (Condon 1997). Systematic errors, however, 
usually limit the actual precision of VLBI astrometry to several times 
this theoretical value (e.g.\
Fomalont et al.\ 1999, Pradel et al.\ 2006). At the frequency of the
present observations, the main sources of systematic errors are
inaccuracies in the troposphere model used, as well as clock, antenna
and {\it a priori} source position errors.  These effects combine to
produce a systematic phase difference between the calibrator and the
target that limits the precision with which the target position can be
determined. We did not attempt to correct for these systematic effects
here, and will, therefore, assume that the true error on each
measurement is the quadratic sum of the random error listed in Tab.\ 1
and a systematic contribution. The latter is difficult to estimate
{\it a priori}, and will be deduced from the fits to the data.\\

\section{Astrometry fits}

The displacement of T Tau Sb on the celestial sphere is the
combination of its trigonometric parallax ($\pi$) and its proper
motion. For isolated sources, it is common to consider linear and
uniform proper motions, so the right ascension ($\alpha$) and the
declination ($\delta$) vary as a function of time $t$ as:

\begin{eqnarray}
\alpha(t) & = & \alpha_0+(\mu_\alpha \cos \delta) t + \pi f_\alpha(t) \label{uni1}\\%
\delta(t) & = & \delta_0+\mu_\delta t + \pi f_\delta(t), \label{uni2}
\end{eqnarray}

\noindent where $\alpha_0$ and $\delta_0$ are the coordinates of the
source at a given reference epoch, $\mu_\alpha$ and $\mu_\delta$ are
the components of the proper motion, and $f_\alpha$ and $f_\delta$ are
the projections over $\alpha$ and $\delta$, respectively, of the
parallactic ellipse. The latter functions are given by (e.g.\
Seidelman 1992):

\begin{eqnarray}
f_\alpha(t) & = & (X \sin \alpha_1 - Y \cos \alpha_1) / (15 \cos \delta_1) \\%
f_\delta(t) & = & (X \cos \alpha_1 \sin \delta_1 + Y \sin \alpha_1 \sin \delta_1 - Z \cos \delta_1),
\end{eqnarray}

\noindent where (X,Y,Z) are the barycentric coordinates of the Earth
in Astronomical Units, and where $\alpha_1$ = $\alpha-\pi f_\alpha(t)$
and $\delta_1$ = $\delta-\pi f_\delta(t)$ are the coordinates of the
barycentric place of the source at each epoch. Note that $f_\alpha$
and $f_\delta$ depend implicitly on time (through X,Y,Z) and
explicitly on the coordinates of the source. The latter dependence on
$\alpha_1$ and $\delta_1$ (which are only known if the trigonometric
parallax is known) implies that the fitting procedure must be
iterative. The barycentric coordinates of the Earth (as well as the
Julian Date of each observation) were calculated using the Multi-year
Interactive Computer Almanac (MICA) distributed as a CDROM by the US
Naval Observatory. They are given explicitly in Tab.\ 2 for all
epochs.
\clearpage
\begin{figure*}[!t]
\centerline{\includegraphics[height=0.8\textwidth,angle=-90]{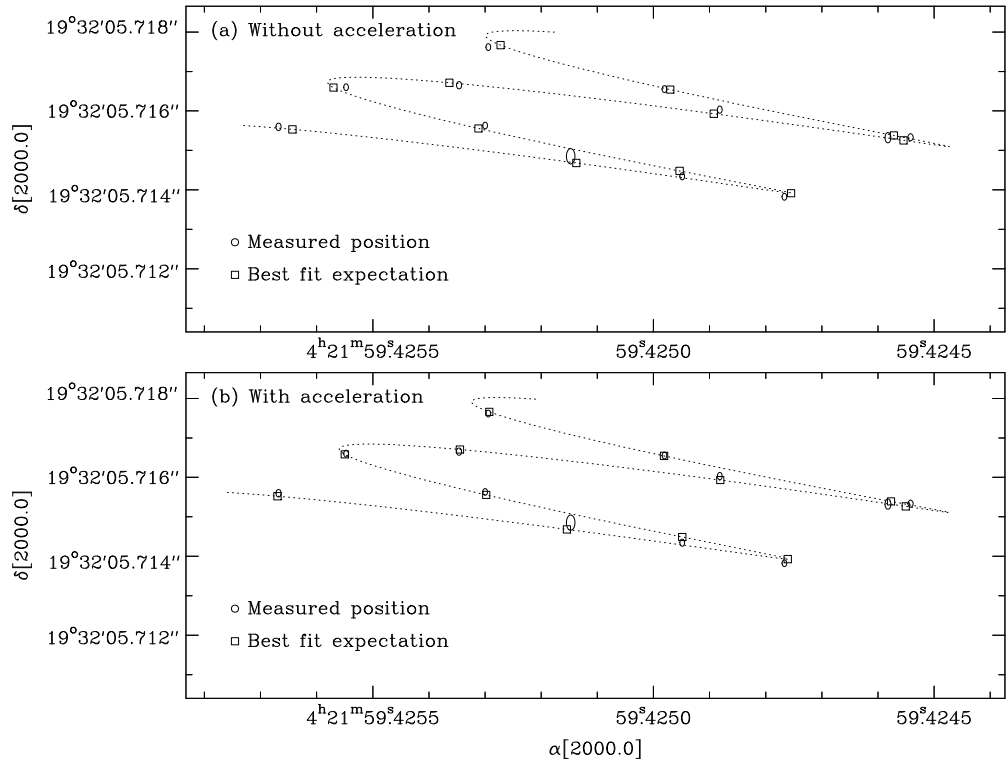}}
\caption{Measured positions of T Tau Sb and best fit without (a) and
with (b) acceleration terms. The observed positions are shown as
ellipses, the size of which represents the magnitude of the
errors. Note the very significant improvement when acceleration terms
are included.}  \end{figure*}
\clearpage

As mentioned earlier, T Tau Sb is a member of a multiple system (e.g.\
Loinard et al.\ 2003, Duch\^ene et al.\ 2006 and references therein),
so its proper motion is likely to be affected by the gravitational
influence of the other members of the system. As a consequence, the
motion is likely to be curved and accelerated, rather than linear and
uniform. To take this into account, we have also made fits to the data
that include acceleration terms. This leads to functions of the form:

\begin{eqnarray}
\alpha(t) & = & \alpha_0+(\mu_{\alpha 0} \cos \delta) t + {1 \over 2} (a_\alpha \cos \delta) t^2 + \pi f_\alpha(t) \label{acc1}\\%
\delta(t) & = & \delta_0+\mu_{\delta 0} t + {1 \over 2} a_\delta t^2 + \pi f_\delta(t), \label{acc2}
\end{eqnarray}

\noindent where $\mu_{\alpha 0}$ and $\mu_{\delta 0}$ are the proper
motions at a reference epoch, and $a_\alpha$ and $a_\delta$ are the
projections of the uniform acceleration. Note that the acceleration
undergone by a body in Keplerian orbit is usually not
uniform. Assuming a uniform acceleration is acceptable here, however,
because our data cover only a small portion ($\sim$ 2 yr) of the
orbital period (a few decades --Duch\^ene et al.\ 2006) of T Tau
Sb. If VLBA data are obtained regularly in the next few decades, a
full orbital fit will become possible, and indeed, necessary.

The astrometric parameters were determined by least-square fitting the
data points with either Eqs.\ \ref{uni1}--\ref{uni2} or Eqs.\
\ref{acc1}--\ref{acc2} using a Singular Value Decomposition (SVD)
scheme (see Appendix for details). To check our results, we also
performed two other fits to the data, a linear one based on the
associated normal equations, and a non-linear one based on the
Levenberg-Marquardt algorithm. They gave results identical to those
obtained using the SVD method. The reference epoch was taken at the
mean of our observations (JD 2453233.586 $\equiv$ J2004.627).

\section{Results}

The fit to the data points by Eqs.\ \ref{uni1}--\ref{uni2} (Fig.\ 
1a) yields the following astrometric parameters:

\begin{eqnarray}
\alpha_{J2004.627} & = & \mbox{ \dechms{04}{21}{59}{425081} } ~ \pm ~ \mbox{ \mmsec{0}{000005} } \nonumber \\%
\delta_{J2004.627} & = & \mbox{ \decdms{19}{32}{05}{71566} } ~ \pm ~ \mbox{ \msec{0}{00003} } \nonumber \\%
\mu_\alpha \cos \delta & = & 4.00 ~ \pm ~ 0.12 ~ \mbox{mas yr$^{-1}$} \nonumber \\%
\mu_\delta & = & -1.18 ~ \pm ~ 0.05 ~ \mbox{mas yr$^{-1}$} \nonumber \\%
\pi & = & 6.90 ~ \pm ~ 0.09 ~ \mbox{mas.} \nonumber
\end{eqnarray}

\noindent
This corresponds to a distance of 145 $\pm$ 2 pc. The post-fit
rms, however, is not very good (particularly in right ascension:
$\sim$ 0.2 mas) as the fit does not pass through many of the observed
positions (Fig.\ 1a).  As a matter of fact, 75 micro-arcsecond and 
--most notably-- 16.5 microseconds of time had to be added 
quadratically to the formal errors listed in Tab.\ 1 to obtain 
a reduced $\chi^2$ of 1 in both right ascension and declination;
the errors on the fitted parameters quoted above include 
this systematic contribution. These large systematic errors most 
certainly reflect the fact mentioned earlier that the proper motion 
of T Tau Sb is not uniform because it belongs to a multiple system. 
Indeed, the fit where acceleration terms are included is significantly 
better (Fig.\ 1b) with a post-fit rms of 60 $\mu$as in right ascension 
and 90 $\mu$as in declination. It yields the following parameters:

\begin{eqnarray}
\alpha_{J2004.627} & = & \mbox{ \dechms{04}{21}{59}{425065} } ~ \pm ~ \mbox{ \mmsec{0}{000002} } \nonumber \\%
\delta_{J2004.627} & = & \mbox{ \decdms{19}{32}{05}{71566} } ~ \pm ~ \mbox{ \msec{0}{0004} } \nonumber \\%
\mu_{\alpha,~J2004.627} \cos \delta & = & 4.02 ~ \pm ~ 0.03 ~ \mbox{mas yr$^{-1}$} \nonumber \\%
\mu_{\delta,~J2004.627} & = & -1.18 ~ \pm ~ 0.05 ~ \mbox{mas yr$^{-1}$} \nonumber \\%
a_\alpha \cos \delta  & = & 1.53 ~ \pm ~ 0.13  ~ \mbox{mas yr$^{-2}$} \nonumber \\%
a_\delta   & = & 0.00 ~ \pm ~ 0.19  ~ \mbox{mas yr$^{-2}$} \nonumber \\%
\pi & = & 6.82 ~ \pm ~ 0.03 ~ \mbox{mas.} \nonumber
\end{eqnarray}

\noindent To obtain a reduced $\chi^2$ of 1 in both right ascension
and declination, one must add quadratically 3.8 microseconds of time
and 75 microseconds of arc to the statistical errors listed in Tab.\
1. The uncertainties reported above and in the rest of this article 
include this systematic contribution. Note also that the reduced 
$\chi^2$ for the fit without acceleration terms is almost 8, if the
latter systematic errors (rather than those mentioned earlier) are used.

The trigonometric parallax obtained when acceleration terms are
included, corresponds to a distance of 146.7 $\pm$ 0.6 pc, somewhat
larger than, but consistent within 1.5$\sigma$ with the value reported
by Loinard et al.\ (2005). Recall, however, that this 2005 result was
based on data that had been corrupted by a problem in the VLBA
correlator; we consider the present value significantly more reliable.
The present distance determination is somewhat smaller than, but
within 1$\sigma$ of the distance obtained by Hipparchos ($d$ =
177$^{+68}_{-39}$ pc). Note that the relative error of our distance is
about 0.4\%, against nearly 30\% for the Hipparchos result, a gain of
almost two orders of magnitude.

\section{Implications for the properties of the stars}

Having obtained an improved distance estimate to the T Tauri system,
we are now in a position to refine the determination of the intrinsic
properties of each of the components of that system. Since the orbital
motion between T Tau N and T Tau S is not yet known to very good
precision, we will use synthetic spectra fitting to obtain the
properties of T Tau N. For the very obscured T Tau S companion, on the
other hand, we will refine the mass determinations based on the
orbital fit obtained by Duch\^ene et al.\ (2006).

\subsection{T Tau N}

The stellar parameters ($T_{\mbox{eff}}$ and $L_{\mbox{bol}}$) of T
Tau N were obtained by fitting synthetic spectra (Lejeune et al.\
1997) to the optical part of the spectral energy distribution. In the
absence of recently published optical spectra with absolute flux
calibration, we decided to use narrow-band photometry taken at six
different epochs from 1965 to 1970 (Kuhi 1974).  In order to eliminate
the contamination by the UV/blue (magnetospheric accretion) and red/IR
(circumstellar disk) excesses, we restricted the fit to the range
0.41--0.65\,$\mu$m.  Two points at $\lambda\lambda$ 0.4340, 0.4861
$\mu$m display large variations between epochs, they were also
discarded as they are likely to be contaminated by emission lines.  As
a consequence, 56 photometric measurements at 13 wavelengths and 6
epochs had to be fitted. (See Fig.\ 2). We assumed that the star kept
constant intrinsic parameters over the 5 years of observation, but
allowed the circumstellar extinction to vary.  Such an hypothesis is
supported by long-term photometric observations (1986-2003) that show
color-magnitude diagrams of T Tau elongated along the extinction
direction (Melnikov \& Grankin 2005); Kuhi (1974) also measured
significant extinction variation in the period 1965-1970 using color
excesses.
\clearpage
\begin{figure}[t]
  \includegraphics[width=\linewidth]{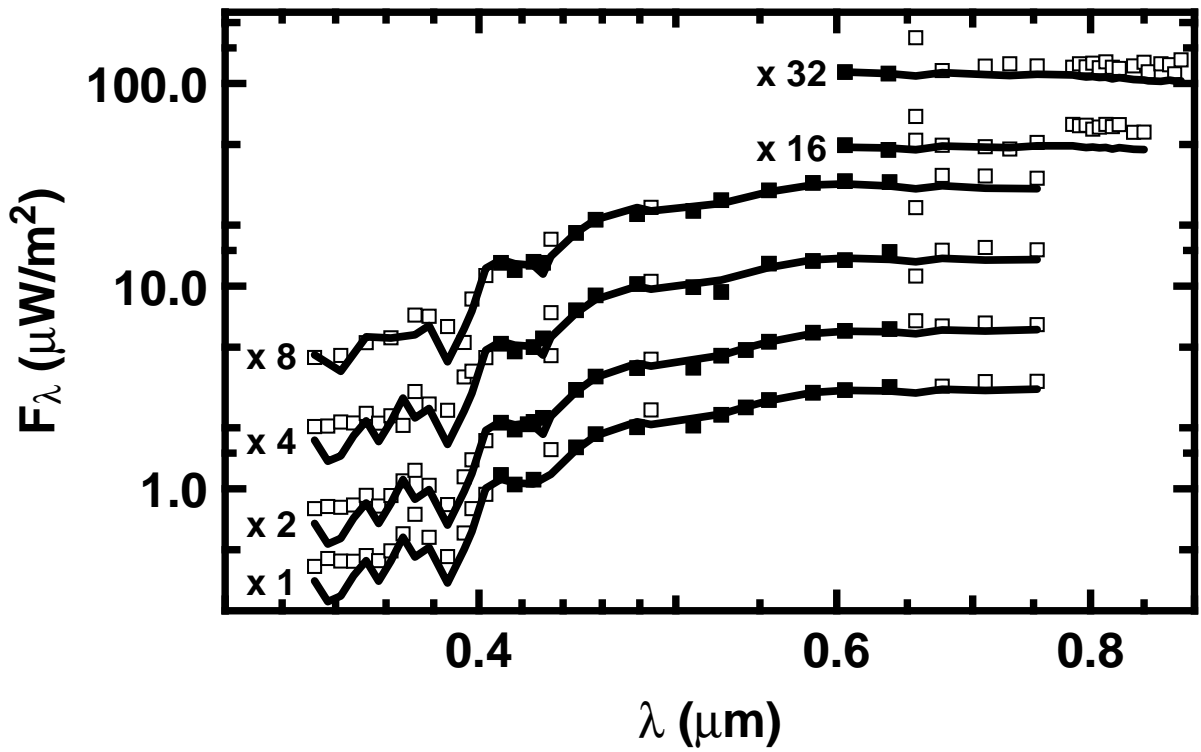}
  \caption{Fit to the photometry at six different epochs.  Black
  squares are point that have been fitted; white squares represent
  other wavelengths excluded from the fit.}
\end{figure}
\clearpage

The non-linear fitting procedure used the Levenberg-Marquardt method
and the determination of errors was done using a Monte Carlo
simulation. The synthetic spectra were transformed into narrow-band
photometry by integration over the bandwidth of the measurements
(typically 0.05\,$\mu$m). As the fitting procedure could not constrain
the metallicity, we assumed a solar one. Several fits using randomly
chosen initial guesses for $T_{\mbox{eff}}$, $L_{\mbox{bol}}$, and
extinctions were performed in order to ensure that a global minimum
$\chi^2$ was indeed reached.  The errors reported by Kuhi (1974;
1.2\%) had to be renormalised to 5.9\% in order to achieve a reduced
$\chi^2$ of 1.  This could result from an underestimation by the
author or from positive and negative contamination by spectral lines
--indeed, Gahm (1970) reports contamination as high as 20\% for RW
Aur.  The best least-squares fit is represented in Fig.\ 3, and yields
$T_{\mbox{eff}} = 5112_{-97}^{+99}$ K and $L_{\mbox{bol}} =
5.11_{-0.66}^{+0.76}\,L_\odot$. The extinction varies between 1.02 and
1.34, within 1-$\sigma$ of the values determined by Kuhi (1974) from
color excesses.  The effective temperature is consistent with a K1
star as reported by Kuhi (1974).

In order to derive the age and mass of T Tau N, pre-main-sequence
isochrones by D'Antona \& Mazzitelli (1996) and Siess et al.\ (2000)
were used. The fitting procedure was identical to the previous one:
the age and mass were converted into effective temperature and
luminosity, which in turn were converted into narrow-band photometry
using the synthetic spectra.  The derived parameters are shown in
Tab.\ 3. The masses (1.83$^{+0.20}_{-0.16}$ and 2.14$^{+0.11}_{-0.10}$
\Msun) have overlapping error bars and are consistent with values
found in the literature (e.g\ Duch\^ene et al.\ 2006). The predicted
ages, on the other hand, differ by a factor of 2. While the
isochrones by D'Antona \& Mazzitelli (1996) give an age in the
commonly accepted range ($1.15^{+0.18}_{-0.16}$ Myr), a somewhat
larger value ($2.39^{+0.31}_{-0.27}$) is derived from Siess et al.\ 
(2000).  Note that the errors on the derived parameters 
are entirely dominated by the modeling errors; the
uncertainty on the distance now represents a very small 
fraction of the error budget.
\clearpage
\begin{figure*}[!t]
\centerline{\includegraphics[height=0.8\textwidth,angle=-90]{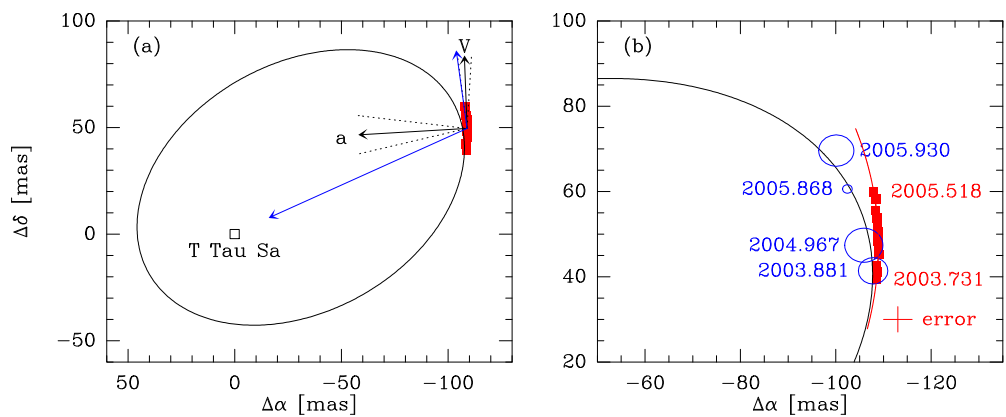}}
\caption{(a) VLBA positions (red squares) registered to T Tau Sa
overimposed on the elliptical fit proposed by Duch\^ene et al.\
(2006). Also shown are the velocity and acceleration vectors for our
mean epoch deduced from our observations, as well as their
counterparts from the fit by Duch\^ene et al.\ (shown in blue). The
dotted black lines around the measured acceleration and velocity
show the error cone on the direction of each of these vectors. (b)
Zoom on the region corresponding to our observations. In addition to
the orbit and the VLBA positions, we show (in red) our best parabolic
fit to our positions, as well as several recent infrared positions (in
blue). The 2003.881 position is from Duch\^ene et al.\ (2005), the 
2004.967 and 2005.868 positions are from Duch\^ene et al.\ (2006),
and the 2005.930 position is from Schaefer et al.\ (2006).}
\end{figure*}
\clearpage

\subsection{T Tau S}

The two members of the T Tau S system have been studied in detail by
Duch\^ene and coworkers in a series of recent articles (Duch\^ene et
al.\ 2002, 2005, 2006). The most massive member of the system (T Tau
Sa) belongs to the mysterious class of ``infrared companions'', and is
presumably the precursor of an intermediate-mass star. T Tau Sb, on
the other hand is a very obscured, but otherwise normal, pre-main
sequence M1 star. The mass of both T Tau Sa and T Tau Sb were
estimated by Duch\^ene et al.\ (2006) using a fit to their orbital
paths. Those authors used the distance to T Tauri deduced from Loinard
et al.\ (2005). Using the new distance determination obtained here, we
can re-normalize those masses. We obtain $M_{Sa}$ = 3.10 $\pm$ 0.34
\Msun, and $M_{Sb}$ = 0.69 $\pm$ 0.18 \Msun. These values may need to
be adjusted somewhat, however, as the fit to the orbital path of the
T Tau Sa/T Tau Sb system is improved (see below). Note finally, 
that the main sources of errors on the masses are related to the 
orbital motion modeling rather than to the uncertainties of 
the distance.

\section{Implications for the orbital motions}

T Tau Sb is a member of a multiple system, so it would be desirable to
give its position and express its motion relative to the other members
of the system, T Tau N and --particularly-- T Tau Sa. Since only T Tau
Sb is detected in our VLBA observations, however, registering the
positions reported here to the other members of the system involves a
number of steps. The absolute position and proper motion of T Tau N
has been measured to great precision using over 20 years of VLA
observations (Loinard et al.\ 2003), so registering the position and
motion of T Tau Sb relative to T Tau N is fairly
straightforward. Combining the data used by Loinard et al.\ (2003)
with several more recent VLA observations, we obtained the following
absolute position (at epoch J2000.0) and proper motion for T Tau N:

\begin{eqnarray}
\alpha_{J2000.0} & = & \mbox{ \dechms{04}{21}{59}{4321} } ~ \pm ~ \mbox{ \mmsec{0}{0001} } \nonumber \\%
\delta_{J2000.0} & = & \mbox{ \decdms{19}{32}{06}{419} } ~ \pm ~ \mbox{ \msec{0}{002} } \nonumber \\%
\mu_{\alpha} \cos \delta & = & 12.35 ~ \pm ~ 0.04 ~ \mbox{mas yr$^{-1}$} \nonumber \\%
\mu_{\delta} & = & -12.80 ~ \pm ~ 0.06 ~ \mbox{mas yr$^{-1}$.} \nonumber
\end{eqnarray}

\noindent
Subtracting these values from the absolute positions and proper motion 
of T Tau Sb, we can obtain the positional offset between T Tau Sb and 
T Tau N, as well as their relative proper motion. For the median epoch 
of our observations, we obtain:

\begin{eqnarray}
\mu_\alpha \cos \delta \mbox{(Sb/N)} & = & -8.33 ~ \pm ~ 0.07 ~ \mbox{mas yr$^{-1}$} \nonumber \\%
\mu_\delta \mbox{(Sb/N)} & = & +11.62 ~ \pm ~ 0.11 ~ \mbox{mas yr$^{-1}$}. \nonumber
\end{eqnarray}

The second step consists in registering the position and motion of T
Tau Sb to the center of mass of T Tau S using the parabolic fits
provided by Duch\^ene et al.\ (2006). Here, both the proper motion and
the acceleration must be taken into account. For the mean epoch of our
observations, we obtain:

\begin{eqnarray}
\mu_\alpha \cos \delta \mbox{(Sb/CM)} & = & +0.3 ~ \pm ~ 0.9 ~ \mbox{mas yr$^{-1}$} \nonumber \\%
\mu_\delta \mbox{(Sb/CM)} & = & +9.3 ~ \pm ~ 0.8 ~ \mbox{mas yr$^{-1}$} \nonumber \\%
a_\alpha \cos \delta \mbox{(Sb/CM)} & = & +1.4 ~ \pm ~ 0.2 ~ \mbox{mas yr$^{-2}$} \nonumber \\%
a_\delta \mbox{(Sb/CM)} & = & -0.1 ~ \pm ~ 0.3 ~ \mbox{mas yr$^{-2}$}. \nonumber 
\end{eqnarray}

The last correction to be made is the registration of the positions, 
proper motions, and accelerations to T Tau Sa rather than to the center 
of mass of T Tau S. This is obtained by simply multiplying the values 
above by the ratio of the total mass of T Tau S (i.e.\ $M_{Sa}$+$M_{Sb}$) 
to the mass of Sa. Using the masses given by Duch\^ene et al.\ 2006), we 
obtain:

\begin{eqnarray}
\mu_\alpha \cos \delta \mbox{(Sb/Sa)} & = & +0.4 ~ \pm ~ 1.1 ~ \mbox{mas yr$^{-1}$} \nonumber \\%
\mu_\delta \mbox{(Sb/Sa)} & = & +11.4 ~ \pm ~ 1.0 ~ \mbox{mas yr$^{-1}$} \nonumber \\%
a_\alpha \cos \delta \mbox{(Sb/Sa)} & = & +1.7 ~ \pm ~ 0.2 ~ \mbox{mas yr$^{-2}$} \nonumber \\%
a_\delta \mbox{(Sb/Sa)} & = & -0.1 ~ \pm ~ 0.3 ~ \mbox{mas yr$^{-2}$}. \nonumber 
\end{eqnarray}

\noindent 
These two vectors are shown in Fig.\ 3 together with the VLBA
positions registered to T Tau Sa, several recent infrared
observations and the elliptical fit obtained by Duch\^ene et al.\
(2006). The final error on the VLBA positions is the combination
of the original uncertainty on their measured absolute position, and 
of the errors made at each of the steps described above. The
final uncertainty is about 3 mas in both right ascension and declination, 
and is shown near the bottom right corner of Fig.\ 3b.

Given the uncertainties, the position of the VLBA source is generally
in good agreement with the infrared source position measured at
similar epochs. Indeed, the first 2 VLBA observations were obtained
almost exactly at the same time as the infrared image published by
Duch\^ene et al.\ (2005), and the positions match exactly. The
position of the VLBA source at the end of 2004 is also in agreement
within 1$\sigma$ with the position of the infrared source at the same
epoch reported by Duch\^ene et al.\ (2006). The situation at the end
of 2005, however, is somewhat less clear. Extrapolating from the last
VLBA observation ($\sim$ 2005.5) to the end of 2005 gives a location
that would be in reasonable agreement with the position given by
Schaefer et al.\ (2006) but clearly not with the position obtained by
Duch\^ene et al.\ (2006). Note, indeed, that the two infrared
positions are only very marginally consistent with one another.

Our VLBA observations suggest that T Tau Sb passed at the westernmost
point of its orbit around 2005.0, whereas according to the fit
proposed by Duch\^ene et al.\ (2006), this westernmost position was
reached slightly before 2004.0. As a consequence, the trajectory
described by the VLBA source is on average almost exactly north-south,
whereas according to the fit proposed by Duch\^ene et al.\ (2006), T
Tau Sb is already moving back toward the east (Fig.\ 3). We note,
however, that the fit proposed by Duch\^ene et al.\ (2006, which gives
an orbital period of 21.7 $\pm$ 0.9 yr) is very strongly constrained by
their 2005.9 observation. Shaefer et al.\ (2006), who measured a
position at the end of 2005 somewhat more to the north (in better
agreement with our VLBA positions), argue that they cannot
discriminate between orbital periods of 20, 30 or 40 yr. Orbits with
longer periods bend back toward the east somewhat later (see Fig.\ 10
in Schaefer et al.\ 2006), and would be in better agreement with our
VLBA positions.

Another element that favors a somewhat longer orbital period is the
acceleration measured here. According to the fit proposed by Duch\^ene
et al.\ (2006), the expected transverse proper motion and acceleration 
are (G.\ Duch\^ene, private communication):

\begin{eqnarray}
\mu_\alpha \cos \delta \mbox{(Sb/Sa)} & = & +1.7 ~ \pm ~ 0.2 ~ \mbox{mas yr$^{-1}$} \nonumber \\%
\mu_\delta \mbox{(Sb/Sa)} & = & +12.1 ~ \pm ~ 1.2 ~ \mbox{mas yr$^{-1}$}. \nonumber \\%
a_\alpha \cos \delta \mbox{(Sb/Sa)} & = & +3.1 ~ \pm ~ 0.5 ~ \mbox{mas yr$^{-2}$} \nonumber \\%
a_\delta \mbox{(Sb/Sa)} & = & -1.4 ~ \pm ~ 0.2 ~ \mbox{mas yr$^{-2}$}. \nonumber
\end{eqnarray}

\noindent Thus, while the expected and observed proper motions are in
good agreement, the expected acceleration is significantly larger that
the observed value (see also Fig.\ 3). A smaller value of the
acceleration would be consistent with a somewhat longer orbital period.

In summary, our observations appear to be in reasonable agreement with
all the published infrared positions obtained over the last few years,
except for the 2005.9 observation reported by Duch\^ene et al.\
(2006).  As a consequence, our data favor an orbital period somewhat
longer than that obtained by Duch\^ene et al.\ (2006). Exactly how
much longer is difficult to assess for the following reason. The orbit
proposed by Duch\^ene et al.\ (2006) was obtained by fitting {\em
simultaneously} the observed positions with a superposition of an
elliptical path (of Sb around Sa) and a parabolic trajectory (of Sa
around N). As a consequence, a modification of the Sa/Sb elliptical
orbit (as may be required by our data) will result in a change in the
parameters of the parabolic fit. But we use the latter to register our
VLBA positions, proper motions, and accelerations against T Tau Sa.
Thus, an entirely new fit will be needed to take into account the
present VLBA observations. Such a fit will be presented in a
forthcoming paper, where the numerous VLA observations available from
the literature, as well as the VLBI observation from Smith et al.\
(2003) will also be taken into account.

\section{Conclusions and perspectives} 

Using a series of 12 radio-continuum VLBA observations of T Tau Sb
obtained roughly every two months between September 2003 and July
2005, we have measured the trigonometric parallax and characterized
the proper motion of this member of the T Tauri multiple system with
unprecedented accuracy. The distance to T Tau Sb was found to be 146.7
$\pm$ 0.6 pc, somewhat larger than the canonical value of 140 pc
traditionally used. Using this precise estimate, we have recalculated
the basic parameters of all three members of the system. The VLBA
positions are in good agreement with recent infrared positions, but
our data seem to favor a somewhat longer orbital period than that
recently reported by Duch\^ene et al.\ (2006) for the T Tau Sa/T Tau Sb
system.

Finally, it should be pointed out that if observations similar to
those presented here were obtained regularly in the coming 5 to 10 
years, they would greatly help to constrain the orbital path (and,
therefore, the mass) of the T Tau Sa/T Tau Sb system.

\acknowledgements
L.L., R.M.T, L.F.R., and R.A.G.\ acknowledge
the financial support of DGAPA, UNAM and CONACyT, M\'exico. NRAO is a
facility of the National Science Foundation operated under cooperative
agreement by Associated Universities, Inc. We are indebted to Gaspard
Duch\^ene for calculating the expected velocity and acceleration from
his fit, and for his comments on the manuscript. We also thank the
anonymous referee for his/her constructive comments on this paper.

\appendix
\section{Appendix}

The parameters determined in this article (position at a reference
epoch, trigonometric parallax, proper motions and accelerations) were
obtained by minimizing the sum ($\chi_\alpha^2+\chi_\delta^2$) of the
residuals in right ascension and declination. The corresponding
general mathematical problem is that where two functions $x$ and $y$
depend linearly on $N$ independent parameters ($a_i$ and $b_i$ for $x$
and $y$, respectively), and $M$ common parameters $c_i$:

\begin{eqnarray}
x(t) & = & \sum_{i = 1}^{N} a_{i} u_{i}(t) + \sum_{j = 1}^{M} c_j w^x_{j}(t)\nonumber\\
y(t) & = & \sum_{i = 1}^{N} b_{i} v_{i}(t) + \sum_{j = 1}^{M} c_j w^y_{j}(t)\nonumber
\end{eqnarray}

\noindent The values $x_k$ and $y_k$ of the functions $x$ and $y$ have
been measured at $P$ times $t_k$ with errors $\sigma^x_k$ and
$\sigma^y_k$, respectively, and the total $\chi^2$ can be written:

\begin{eqnarray}
\chi^2 & = & \chi_x^2+\chi_y^2 \nonumber\\
       & = & \sum_{k=1}^{P} \left( \left[ {x_k- \left( \sum_{i = 1}^{N} a_{i} u_{i}(t_k) + \sum_{j = 1}^{M} c_j w^x_{j}(t_k) \right) \over \sigma^x_k} \right]^2 + \left[ {y_k- \left( \sum_{i = 1}^{N} b_{i} v_{i}(t_k) + \sum_{j = 1}^{M} c_j w^y_{j}(t_k) \right) \over \sigma^y_k} \right]^2 \right)\nonumber\\ \label{chi2}
\end{eqnarray}

\noindent Defining the following matrix elements:

\begin{eqnarray}
\alpha_{ki} & = & {u_i(t_k) \over \sigma_k^x} \nonumber \\
\beta_{ki}  & = & {v_i(t_k) \over \sigma_k^y} \nonumber \\
\gamma_{kj} & = & {w_j^x(t_k) \over \sigma_k^x} \nonumber \\
\delta_{kj} & = & {w_j^y(t_k) \over \sigma_k^y} \nonumber \\
\theta_k    & = & {x_k \over \sigma_k^x} \nonumber \\
\psi_k      & = & {y_k \over \sigma_k^y}, \nonumber
\end{eqnarray}

\noindent we can re-write the total $\chi^2$ given by equation
\ref{chi2} as:

\begin{eqnarray}
\chi^2 & = & \sum_{k=1}^{P} \left( \left[ \theta_k - \sum_{i=1}^{N} \alpha_{ki}a_i - \sum_{j=1}^{M} \gamma_{kj}c_j \right]^2 + \left[ \psi_k - \sum_{i=1}^{N} \beta_{ki}b_i - \sum_{j=1}^{M} \delta_{kj}c_j \right]^2 \right) \nonumber
\end{eqnarray}

\noindent
This sum of quadratic terms can clearly be seen as the squared 
norm of the vector:

\[ \left( \begin{array}{c} \theta_1 - \sum_{i=1}^{N} \alpha_{1i}a_i - \sum_{j=1}^{M} \gamma_{1j}c_j\\
                           \theta_2 - \sum_{i=1}^{N} \alpha_{2i}a_i - \sum_{j=1}^{M} \gamma_{2j}c_j\\
                           \dotfill\\
                           \theta_P - \sum_{i=1}^{N} \alpha_{Pi}a_i - \sum_{j=1}^{M} \gamma_{Pj}c_j\\
                           \psi_1 - \sum_{i=1}^{N} \beta_{1i}b_i - \sum_{j=1}^{M} \delta_{1j}c_j\\
                           \psi_2 - \sum_{i=1}^{N} \beta_{2i}b_i - \sum_{j=1}^{M} \delta_{2j}c_j\\
                           \dotfill\\
                           \psi_P - \sum_{i=1}^{N} \beta_{Pi}b_i - \sum_{j=1}^{M} \delta_{Pj}c_j\\
          \end{array} \right) \]

\noindent
Re-arranging the terms in this expression, we can re-write the total $\chi^2$ as:

\begin{equation}
\chi^2 = \left|\left| \left( \begin{array}{ccc} \alpha & 0 & \gamma \\
                                              0 & \beta & \delta \end{array} 
                                \right) \left( \begin{array}{c} a \\ b \\ c \end{array} \right) - \left( \begin{array}{c} \theta\\ \psi \end{array} \right) \right|\right|^2 \label{altchi2}
\end{equation}

\noindent The procedure then consists of finding the vector $\mathbf X$ 
that minimizes an expression of the form:

\[ \left| \left| A.X - B \right| \right|^2 \]

\noindent An efficient algorithm to perform this operation is known as
the {\em Singular Value Decomposition} (SVD --see Press et al.\
1992). This method is based on the linear algebra theorem that states
that any $I \times J$ rectangular matrix $\mathbf M$ whose number $I$
of rows is larger or equal to its number $J$ of columns, can written
as the product of an $I \times J$ column-orthogonal matrix $\mathbf U$
by a $J \times J$ diagonal matrix $\mathbf W$ with positive or zero
elements, and the transpose of a $J \times J$ orthogonal matrix
$\mathbf V$:

\begin{equation} M = U ~W ~V^T \label{decomp} \end{equation}

\noindent The rectangular matrix in expression \ref{altchi2} has $2P$
rows (=24 in our case) and $2N+M$ column (=5 or 7, for uniform or
accelerated proper motions, respectively) and can clearly be
decomposed in that fashion. Since both $\mathbf U$ and $\mathbf V$ in
the previous expression are orthogonal, their inverses are just their
transposes. Also, if none of its diagonal elements are zero (which
will be the case in all situations considered here), the inverse of
$\mathbf W$ is a diagonal matrix whose elements are just the inverses
of those of $\mathbf W$. Thus, the inverse of matrix $\mathbf M$ can
be written as:

\[ M^{-1} = V ~W^{-1} ~U^T \]

\noindent
It can be shown (see Press et al.\ 1992) that, if the matrix $\mathbf M$ 
can be decomposed as above, then the vector $\mathbf X$ that minimizes
the expression $\left| \left| A.X - B \right| \right|^2$ is simply:

\begin{equation} X = V ~W^{-1} ~U^T B \label{sol} \end{equation}

\noindent An efficient way of solving our least-squares fit problem is,
therefore, to form the rectangular matrix that appears in expression
\ref{altchi2}, decompose it as in \ref{decomp}, and calculate the
value of the $a_i$'s, $b_i$'s, and $c_i$'s using \ref{sol}. This
method was implemented in FORTRAN following Press et al.\ (1992).

\begin{deluxetable}{llllllr}
\rotate
\tablewidth{0pt}
\tablecaption{Source position and flux}
\tablehead{
\colhead{Mean UT date}    &  
\colhead{$\alpha$ (J2000.0)} &
\colhead{$\sigma_\alpha$} &
\colhead{$\delta$ (J2000.0)} &
\colhead{$\sigma_\delta$} &
\colhead{$F_\nu$} &
\colhead{$\sigma$}\\%
~~(yyyy.mm.dd ~~ hh:mm)~~ & & & & & (mJy) & ($\mu$Jy)}
\startdata
2003.09.24 ~~ 11:33 \dotfill & \dechms{04}{21}{59}{4252942} & \mmsec{0}{0000013} & \decdms{19}{32}{05}{717618} & \msec{0}{000043} & 1.62 & 74 \\%
2003.11.18 ~~ 08:02 \dotfill & \dechms{04}{21}{59}{4249805} & \mmsec{0}{0000015} & \decdms{19}{32}{05}{716554} & \msec{0}{000043} & 1.74 & 66 \\%
2004.01.15 ~~ 04:09 \dotfill & \dechms{04}{21}{59}{4245823} & \mmsec{0}{0000036} & \decdms{19}{32}{05}{715322} & \msec{0}{000108} & 0.92 & 72 \\%
2004.03.26 ~~ 23:26 \dotfill & \dechms{04}{21}{59}{4245420} & \mmsec{0}{0000017} & \decdms{19}{32}{05}{715333} & \msec{0}{000050} & 1.27 & 70  \\%
2004.05.13 ~~ 20:17 \dotfill & \dechms{04}{21}{59}{4248818} & \mmsec{0}{0000016} & \decdms{19}{32}{05}{716034} & \msec{0}{000055} & 1.90 & 111 \\%
2004.07.08 ~~ 16:37 \dotfill & \dechms{04}{21}{59}{4253464} & \mmsec{0}{0000020} & \decdms{19}{32}{05}{716652} & \msec{0}{000058} & 1.25 & 64  \\%
2004.09.16 ~~ 11:59 \dotfill & \dechms{04}{21}{59}{4255476} & \mmsec{0}{0000015} & \decdms{19}{32}{05}{716602} & \msec{0}{000042} & 1.61 & 70 \\%
2004.11.09 ~~ 08:27 \dotfill & \dechms{04}{21}{59}{4252999} & \mmsec{0}{0000015} & \decdms{19}{32}{05}{715631} & \msec{0}{000039} & 3.36 & 104 \\%
2004.12.28 ~~ 05:14 \dotfill & \dechms{04}{21}{59}{4249488} & \mmsec{0}{0000015} & \decdms{19}{32}{05}{714344} & \msec{0}{000050} & 1.26 & 70 \\%
2005.02.24 ~~ 01:26 \dotfill & \dechms{04}{21}{59}{4247667} & \mmsec{0}{0000016} & \decdms{19}{32}{05}{713826} & \msec{0}{000042} & 2.30 & 80 \\%
2005.05.09 ~~ 20:32 \dotfill & \dechms{04}{21}{59}{4251475} & \mmsec{0}{0000060} & \decdms{19}{32}{05}{714852} & \msec{0}{000182} & 1.12 & 106 \\%
2005.07.08 ~~ 16:36 \dotfill & \dechms{04}{21}{59}{4256679} & \mmsec{0}{0000019} & \decdms{19}{32}{05}{715598} & \msec{0}{000060} & 1.41 & 75 
\enddata
\end{deluxetable}

\begin{deluxetable}{lllll}
\tablecaption{Julian dates and Earth coordinates}
\tablehead{
\colhead{Mean UT date}    &
\colhead{JD}      &
\multicolumn{3}{c}{Earth Barycentric coordinates} \\%
(yyyy.mm.dd ~~ hh.mm) & & \multicolumn{3}{c}{Astronomical Units}}
\startdata
2003.09.24 ~~ 11:33 \dotfill & 2452906.981522 & $+$1.006064570 & $+$0.012414145 & $+$0.005329883 \\%
2003.11.18 ~~ 08:02 \dotfill & 2452961.834705 & $+$0.563031813 & $+$0.744728145 & $+$0.322808936 \\%
2004.01.15 ~~ 04:09 \dotfill & 2453019.672980 & $-$0.401044530 & $+$0.820090543 & $+$0.355473794 \\%
2004.03.26 ~~ 23:26 \dotfill & 2453091.476395 & $-$0.987542097 & $-$0.107099184 & $-$0.046513144 \\%
2004.05.13 ~~ 20:17 \dotfill & 2453139.345324 & $-$0.599629779 & $-$0.745571223 & $-$0.323319720 \\%
2004.07.08 ~~ 16:37 \dotfill & 2453195.192419 & $+$0.297481514 & $-$0.894474582 & $-$0.387883413 \\%
2004.09.16 ~~ 11:59 \dotfill & 2453264.999583 & $+$1.003677539 & $-$0.099008512 & $-$0.043027999 \\%
2004.11.09 ~~ 08:27 \dotfill & 2453318.852141 & $+$0.676914299 & $+$0.666368259 & $+$0.288787192 \\%
2004.12.28 ~~ 05:14 \dotfill & 2453367.718351 & $-$0.111308860 & $+$0.895710071 & $+$0.388210192 \\%
2005.02.24 ~~ 01:26 \dotfill & 2453425.559664 & $-$0.896015668 & $+$0.377089952 & $+$0.163364585 \\%
2005.05.09 ~~ 20:32 \dotfill & 2453500.355214 & $-$0.655044089 & $-$0.701050689 & $-$0.304055855 \\%
2005.07.08 ~~ 16:36 \dotfill & 2453560.191348 & $+$0.293538593 & $-$0.893249243 & $-$0.387385193
\enddata
\end{deluxetable}

\begin{table}[t]
  \caption{Parameters of T~Tauri N}
  \label{tab:photo-fit}
  \begin{tabular}{ccccc}
  \hline\hline
  Isochrone set & Siess et al.\ (2000) & D'Antona \& Mazzitelli (1997)\\
  \hline
  Age (Myr)        & $2.39^{+0.31}_{-0.27}$        & $1.15^{+0.18}_{-0.16}$\\
  Mass ($M_\odot$) & $2.14^{+0.11}_{-0.10}$        & $1.83^{+0.20}_{-0.16}$\\
  \hline
  $T_{\mbox{eff}}$ (K)         & \multicolumn{2}{c}{$5112^{+99}_{-97}$}\\
  $L_{\mbox{bol}}$ ($L_\odot$) & \multicolumn{2}{c}{$5.11^{+0.76}_{-0.66}$}\\
  $R_{\star}$ ($R_\odot$)      & \multicolumn{2}{c}{$2.89^{+0.24}_{-0.21}$}\\
  ${A_V}_{\mbox{MJD 39095.2}}$ & \multicolumn{2}{c}{$1.34\pm0.17$}\\
  ${A_V}_{\mbox{MJD 39153.2}}$ & \multicolumn{2}{c}{$1.37\pm0.17$}\\
  ${A_V}_{\mbox{MJD 39476.3}}$ & \multicolumn{2}{c}{$1.20\pm0.17$}\\
  ${A_V}_{\mbox{MJD 40869.4}}$ & \multicolumn{2}{c}{$1.02\pm0.17$}\\
  ${A_V}_{\mbox{MJD 39485.1}}$ & \multicolumn{2}{c}{$1.36\pm0.19$}\\
  ${A_V}_{\mbox{MJD 39524.1}}$ & \multicolumn{2}{c}{$1.16\pm0.19$}\\
  \hline\\
  \end{tabular}
\end{table}

\end{document}